\RequirePackage{fix-cm}
\documentclass[twocolumn]{svjour3}          
\smartqed  
\usepackage{graphicx}
\usepackage{algorithm}
\usepackage[noend]{algpseudocode}
\usepackage{amsmath}
\usepackage{amssymb}
\usepackage{subfigure}
\usepackage{multirow}
\begin{document}

\title{Feature extraction without learning in an analog Spatial Pooler memristive-CMOS circuit design of Hierarchical Temporal Memory  
}

\titlerunning{HTM feature extraction without learning}        

\author{Olga Krestinskaya         \and
        Alex Pappachen James 
}


\institute{O. Krestinakaya and A.P.James\at
              Nazarbayev University, Astana, Kazakhstan \\
              \email{okrestinskaya@nu.edu.kz;apj@ieee.org} \\
              http://www.biomicrosystems.info/alex
}

\date{Received: date / Accepted: date}

\maketitle

\begin{abstract}
Hierarchical Temporal Memory (HTM) is a neuromorphic algorithm that emulates sparsity, hierarchy and modularity resembling the working principles of neocortex. Feature encoding is an important step to create sparse binary patterns. This sparsity is introduced by the binary weights and random weight assignment in the initialization stage of the HTM. We propose the alternative deterministic method for the HTM initialization stage, which connects the HTM weights to the input data and preserves natural sparsity of the input information. Further, we introduce the hardware implementation of the deterministic approach and compare it to the traditional HTM and existing hardware implementation. We test the proposed approach on the face recognition problem and show that it outperforms the conventional HTM approach.

\end{abstract}

\section{Introduction}
\label{intro}

Hierarchical Temporal Memory (HTM) is  a neuromorphic algorithm that emulates the structure and functionality of the cortical neural networks \cite{fedorova2016htm}. HTM can serve as a tool for intelligent data processing in edge computing devices. The increase in the number of edge computing devices and Internet of things (IoT) applications in the recent years lead to the demand to introduce on sensor processing using analog hardware. Therefore, the translation of the HTM algorithm into analog hardware can produce the promising solution to the computational speed problems  \cite{csapo2007object,ibrayevdesign,tcad}.

HTM is divided into two parts: HTM Spatial Pooler (HTM SP) and HTM Temporal Memory (HTM TM). The HTM SP has been proven to be useful for visual data processing and classification problems, whereas the HTM TM is used as a prediction and learning algorithm. In this work, we focus on the SP part of HTM. The main functionality of the HTM SP is to form the sparse distributed pattern from the input data and perform the feature encoding. The recent works show that it is useful for feature extraction and pattern recognition problems \cite{tcad}.

In this work, we investigate the initialization stage of the HTM SP and proposed the rule-based deterministic approach instead of the random weight approach for the initial weight assignment. The main purpose of the rule-based approach is to connect the input to the HTM weights, which allows to preserve natural sparsity and structural information from the inputs. Moreover, we propose the hardware implementation for the rule-based approach and compare it with the conventional random weight approach in terms of power dissipation and on-chip area requirements. Also, we test the system level implementation of the proposed approach on the face recognition problem and show the improvements in the  recognition accuracy \cite{james2017design,tcad}.  

This paper is organized into 8 sections. Section \ref{s2} provides the overview of the HTM algorithm and introduces the mathematical framework of HTM. Section \ref{s3} illustrates the difference between the conventional approach and the proposed rule-based approach. Section \ref{s4} discussed the hardware implementation of the HTM SP and illustrates the proposed hardware architecture. Section \ref{s5} shows how system level HTM SP algorithm can be used for the face recognition problem. Section \ref{s6} shows the results of the system-level and analog hardware implementations. Section \ref{s7} provided the discussion of the proposed rule-based method and corresponding analog hardware. Section \ref{s8} concludes the paper.

\section{Background}
\label{s2}
\subsection{HTM overview}

HTM is a neuromorphic machine learning algorithm that emulates the architecture and biological functionality of the neocortex in a human brain \cite{hawkinsintelligence}. 
HTM algorithm focuses on the sparse distributed representation of the information, encoding of the input sensory data, learning and prediction making based on the temporal changes in the input data and previous inputs \cite{george2005hierarchical}.

As discussed in the introduction section, the original HTM algorithm is divided into two main parts: Spatial Pooler (SP) and Temporal Memory (TM). The main purpose of the HTM SP is the encoding and producing sparse distributed representation (SDR) of the input data. This is useful for the feature extraction and visual data classification purposes  \cite{zyarah2015design}. The applications of the HTM SP include handwritten digits recognition \cite{fan2016hierarchical}, face recognition \cite{tcad}, speech recognition  \cite{fedorova2016htm}, gender classification \cite{james2017design}, object categorization \cite{csapo2007object} and natural language processing \cite{poster}. The HTM TM is responsible for the learning and processing of temporal patterns and can be used for the prediction taking into account previous experiences \cite{tcad}.

\begin{figure}
\centering
\includegraphics[width=80mm]{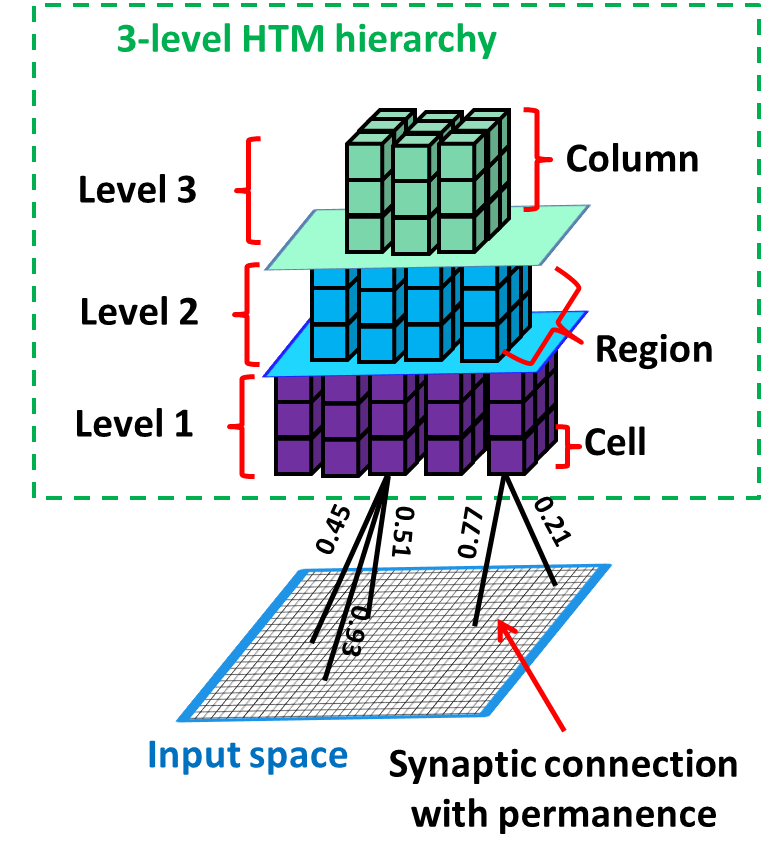}
\caption{The hierarchical structure of HTM. The example shows 3-level HTM, where each level consists of the regions with columns, and columns are created from the cells. HTM is connected to the input space by the synapses.}
\label{fig1}
\end{figure}

HTM has a hierarchical structure, and an example 3-level HTM structure is shown in Fig. \ref{fig1}. Each level of HTM consists of certain regions with columns, and each column is comprised of cells. The columns in HTM is equivalent to neurons. The columns are connected to the input space with the connections through the dendrite segment with several synapses. Each synapse has a certain weight called synaptic permanence.

The HTM SP consists of four main phases: initialization, calculation of overlap value, inhibition and learning process \cite{hawkinsintelligence}. In the initialization phase, the potential inputs in the HTM regions are identified and certain number of columns within the HTM regions are selected \cite{spnew}. The potential input columns are those columns, which are considered to receive the input data. The inputs are connected to the potential input columns through the dendrite containing with potential synapses with certain permanence value (weight). In the initialization phase, the weights are assigned randomly following the uniform random distribution approach. If this weight (permanence value) is greater than the threshold, the potential synapse is considered to be connected. If the connected synapse is connected to the active input it is considered to be active connected. In the overlap phase the number of active connected synapses is computed. In the inhibition stage, the $k$ columns with highest overlap values become active (assigned as high, 1), and the other columns are inhibited (assigned as low, 0). In the learning phase, the HTM SP weights of the synapses are updated based on the Hebb`s leaning rules. After the update process, all phases, except initialization phase, are repeated.

\subsection{Mathematical framework of the HTM SP}

\begin{figure}
\centering
\includegraphics[width=80mm]{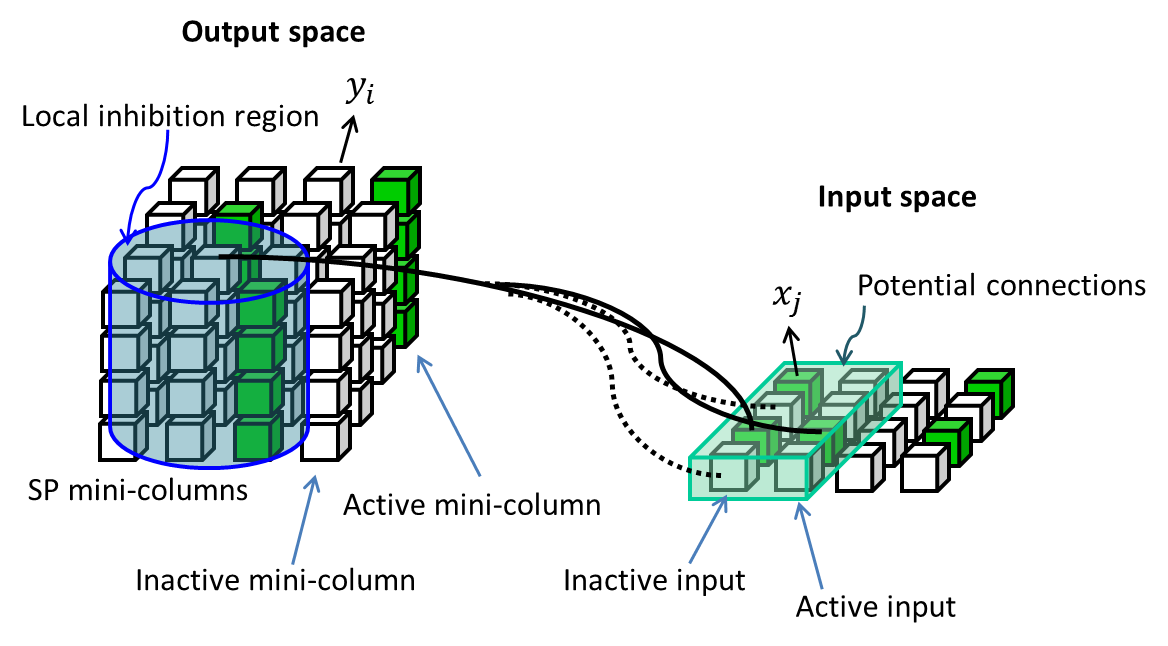}
\caption{The arrangement of the HTM SP input space and the output space, containing the mini-columns.}
\label{fig2}
\end{figure}

The arrangement of the input of the HTM SP and output space that is arranged into mini-columns is shown in Fig.\ref{fig2}. The parameter $x_j$ denotes the $j$-th input neuron in the input space, and the $y_i$ refers to the $i$th output SP mini-column in the output space, which is connected to the region of the input space with the potential connections. 

The synapse of the $i$-th SP mini-column is located in a hypercube of the input space centered at $x_i^c$ with the edge length of $\gamma$. The potential connections are defined in Eq.\ref{olgaeq1}, where $\imath(x_j;x_i^c,\gamma)=1, \forall x_j\in (x_i^c,\gamma)$, and $z_{ij}\sim U(0,1)$, $z$ is selected randomly and follows the uniform distribution rule ($U$ has a range: $[0,1]$) \cite{25,tcad}. The parameter $\rho$ denotes the assigned percentage of inputs that are considered to be potential connections within the hypercube of the input space.
\begin{equation}
PI(i)= \left \{ j|\imath(x_j;x_i^c,\gamma)  \;\textup{and} \; (z_{ij}<\rho) \right \}
\label{olgaeq1}
\end{equation}

 For all synapses the synaptic permanence value (weight) is assigned. The synaptic permanence from the $j$-th input to the $i$-th SP mini-column is represented by the matrix $\mathbf{S_{ij}} \in [0,1]$ shown in Eq.\ref{olgaeq3}. If the synapse is located within the potential the region of potential inputs, the synaptic permanence value $\mathbf{S}_{ij}$ is assigned as the uniform random distribution between 0 and 1; otherwise, the synaptic permanence is 0, so the synapse is not connected. 

\begin{equation}
\mathbf{S}_{ij}= \left\{\begin{matrix}
 U(0,1)& \textup{if} \quad j\in PI(i)\\ 
 0& \textup{otherwise}
\end{matrix}\right.
\label{olgaeq3}
\end{equation}

All the connected synapses are represented by the binary matrix $\mathbf{B}$ shown in Eq.\ref{eqolga2}. Based on the synaptic permanence value, the synapse is either connected or not connected. If their value is greater the threshold value $\theta_c$, the synapse is connected and $\mathbf{B}=1$, and vise versa. The threshold $\theta_c$ shows the percentage of the connected synapses.

\begin{equation}
\mathbf{B}_{ij}= \left\{\begin{matrix}
 1& \textup{if} \quad \mathbf{S_{ij}}\geq \theta_c \\ 
 0& \textup{otherwise}
\end{matrix}\right.
\label{eqolga2}
\end{equation}

Equation \ref{olgaeq4} refers to the process, where the local inhibition neighborhood region $\mathbf{N}_i$ of the HTM SP of the $i$-th SP mini-column is determined. The parameter $\left \| y_i-y_j \right \|$ refers to the the Euclidean distance between the mini-columns $i$ and $j$, and the parameter $\phi$ controls the inhibition radius.

\begin{equation}
\mathbf{N}_i= \left\{ j|\left \| y_i-y_j \right \|<\phi, i\ne j \right\}
\label{olgaeq4}
\end{equation}

In the overlap phase of the HTM SP, the activation of the SP mini-columns for a particular input pattern $\mathbf{Z}$ is determined. The input overlap calculation is shown in Eq.\ref{olgaeq5}, where  $\beta_i$ is a boosting factor that refers to the excitability of the SP mini-column.

\begin{equation}
o_i= \beta_i \sum_j \mathbf{B}_{ij}\mathbf{Z}_j
\label{olgaeq5}
\end{equation}

In the inhibition phase, the activation of the SP mini-columns occurs. The activation depends of two conditions: the value of the input overlap of the SP mini-column should be above the threshold $\theta_s$ and within the top $s$ percent considering the other SP mini-columns in the inhibition neighborhood. The selection of the active column is shown in Eq.\ref{olgaeq6}, where the parameter $\alpha_i$ is the activity of the SP mini-columns, $prctile$ is a percentile function, and $\mathbf{NO}(i)=\left\{o_j|j\in \mathbf{N}(i) \right\}$ with the target activation density $s$. The activation of the columns is implemented according to the $k$-winners-take-all rule considering all mini-columns in the particular neighborhood.

\begin{equation}
\alpha_i= 1, \textup{if} \quad (o_i \geq \textup{prctile}(\mathbf{NO}(i),1-s)) \;\textup{and}\; (o_i\geq \theta_s)
\label{olgaeq6}
\end{equation}

 In the original HTM algorithm, the parameter $k$, can be changed based on the desired number of winning columns for a particular application \cite{25}. However, in most of the existing hardware implementations of the HTM SP, $k=1$ due to the limitations of the Winner-Takes-All(WTA) circuits \cite{ibrayevdesign}.

In the learning phase of the HTM SP, feed-forward connections are learned using Hebb`s learning rule and the boosting factor is updated. The Hebb`s rule for the connection learning implies that the permanence value of the connections is either increased or decreased by the value $\rho$. The update process of the boosting factor is performed considering time-average activity the SP mini-columns $\bar{\alpha_i} (t)$ and recent activity of the SP mini-columns $<\bar{\alpha_i} (t)>$ \cite{25}. Eq.\ref{olgaeq66} shows the calculation of the time-average activity of the SP mini-columns in time $t$, where $T$ is the number of considered previous inputs, and $\alpha_i (t)$ is a current activity the $i$-th mini-column.

\begin{equation}
\bar{\alpha_i} (t)= \frac{(T-1)\times \bar{\alpha_i}(t-1)+\alpha_i (t)}{T}
\label{olgaeq66}
\end{equation}

Equation \ref{olgaeq7} shows the calculation of the recent activity.

\begin{equation}
<\bar{\alpha_i} (t)>= \frac{1}{|\mathbf{N}(i)|} \displaystyle\sum_{j\in \mathbf{N}(i)}^{} \bar{\alpha_i} (t)
\label{olgaeq7}
\end{equation}

Equation \ref{olgaeq8} refers to the update process of the boosting factor, where $\eta$ controls the adaptation of the HTM SP.

\begin{equation}
\beta_i (t)= e^{-\eta(\bar{\alpha_i} (t)-<\bar{\alpha_i} (t)>)}
\label{olgaeq8}
\end{equation}

\section{Rule-based approach}
\label{s3}
To improve the initialization phase of the HTM SP, we proposed the rule-based approach for the weights assignment instead of the uniform weight distribution. In the rule-based approach, we establish the connection between the input space and the synaptic permanence values (weights of the synapses). The Eq. \ref{o10} shows how synaptic permanence weights are assigned in the rule based approach. Eq. \ref{o10} is used instead of Eq. \ref{olgaeq3} and Eq. \ref{eqolga2}. 

\begin{equation}
\mathbf{S}_{ij}= \left\{\begin{matrix}
 1& \textup{if} \quad j\in PI(i) \: \textup{and} \: PI(i)>mean(PI)\\ 
 0& \textup{otherwise}
\end{matrix}\right.
\label{o10}
\end{equation}

In the rule-based approach, the synaptic permanence value is assigned based on the mean value of the inputs within the input space region with the potential connections. If the input is greater than the mean of the the inputs within this neighborhood, the synaptic permanence $\mathbf{S}_{ij}$ is 1, otherwise $\mathbf{S}_{ij}=0$. 

In this work, we focus on the first three phases of the HTM SP: initialization, overlap and inhibition. The Algorithm \ref{alg} summaries the proposed approach.  Lines 2-18 represent the HTM SP initialization stage, lines 20-22 refer to the overlap stage, and lines 24-27 correspond to the inhibition stage of the HTM SP.

\begin{algorithm}[t]
\caption{The HTM SP algorithm}\label{alg}
\begin{algorithmic}[1]

\State {} 
\Comment{\textbf{HTM SP initialization}}
\State {Define the size of input neighborhood with potential connections, $x_i^c$, $\gamma$, $\rho$, $\eta$, $\theta_c$, size of the local inhibition region, $\theta_s$}
\State{Determine $\phi$ by multiplying the average number of connected input spans of all the SP mini-columns by the number of mini-columns per inputs.}

 \State $z_{ij}\sim U(0,1)$

 \If{$\forall x_j\in (x_i^c,\gamma)$} 
 \State $\imath(x_j;x_i^c,\gamma)=1$
 \EndIf

 \For{$\imath(x_j;x_i^c,\gamma)  \;\textup{and} \; (z_{ij}<\rho)$} 
 \State $PI(i)= j$
 \EndFor

 \If{$j\in PI(i)\: \textup{and} \: PI(i)>mean(PI)$} 
 \State $\mathbf{S}_{ij}= 1$
 \Else
 \State $\mathbf{S}_{ij}=0$
 \EndIf

 \State $\mathbf{B}_{ij}=\mathbf{S}_{ij}$

 \For{$| y_i-y_j |<\phi, i\ne j $} 
 \State $\mathbf{N}_i=  j$
 \EndFor
\State {} 
\Comment{\textbf{HTM SP oveplap}}
\State{$o_i= \beta_i \mathbf{B}_{ij}\mathbf{Z}_j$}
 \For{$j\in \mathbf{N}(i) $} 
 \State $\mathbf{NO}(i)=o_j$
 \EndFor
 
 \State {} 
\Comment{\textbf{HTM SP inhibition}}
 \If{$(o_i \geq \textup{prctile}(\mathbf{NO}(i),1-s)) \;\textup{and}\; (o_i\geq \theta_s)$} 
 \State $\alpha_i= 1$
 \Else
 \State $\alpha_i= 0$
 \EndIf
 
\end{algorithmic}
\end{algorithm}

\section{Hardware implementation}
\label{s4}
\subsection{Modified HTM SP}
In this work, we investigate the modified HTM approach proposed in \cite{tcad}. The difference between the original algorithm and the modified version of HTM is in the activation of the columns in the inhibition stage. The inhibition stage of the original algorithm is based on the   WTA approach of $k$ largest overlap values. In the modified version of the HTM SP, the selection of the winning columns occurs based on the mean value of the overlap in the inhibition region. If the overlap value of the column is greater than the mean value of the overlaps in the inhibition region, the columns is activated, otherwise, it is inhibited. The modified approach for the inhibition region is represented in Eq. \ref{olgaeq11}, which is used instead of the Eq. \ref{olgaeq6}.

\begin{equation}
\alpha_i= 1, \textup{if} \quad (o_i \geq mean(o_j)|j\in \mathbf{N}(i))
\label{olgaeq11}
\end{equation}

\begin{figure}
\centering
\includegraphics[width=80mm]{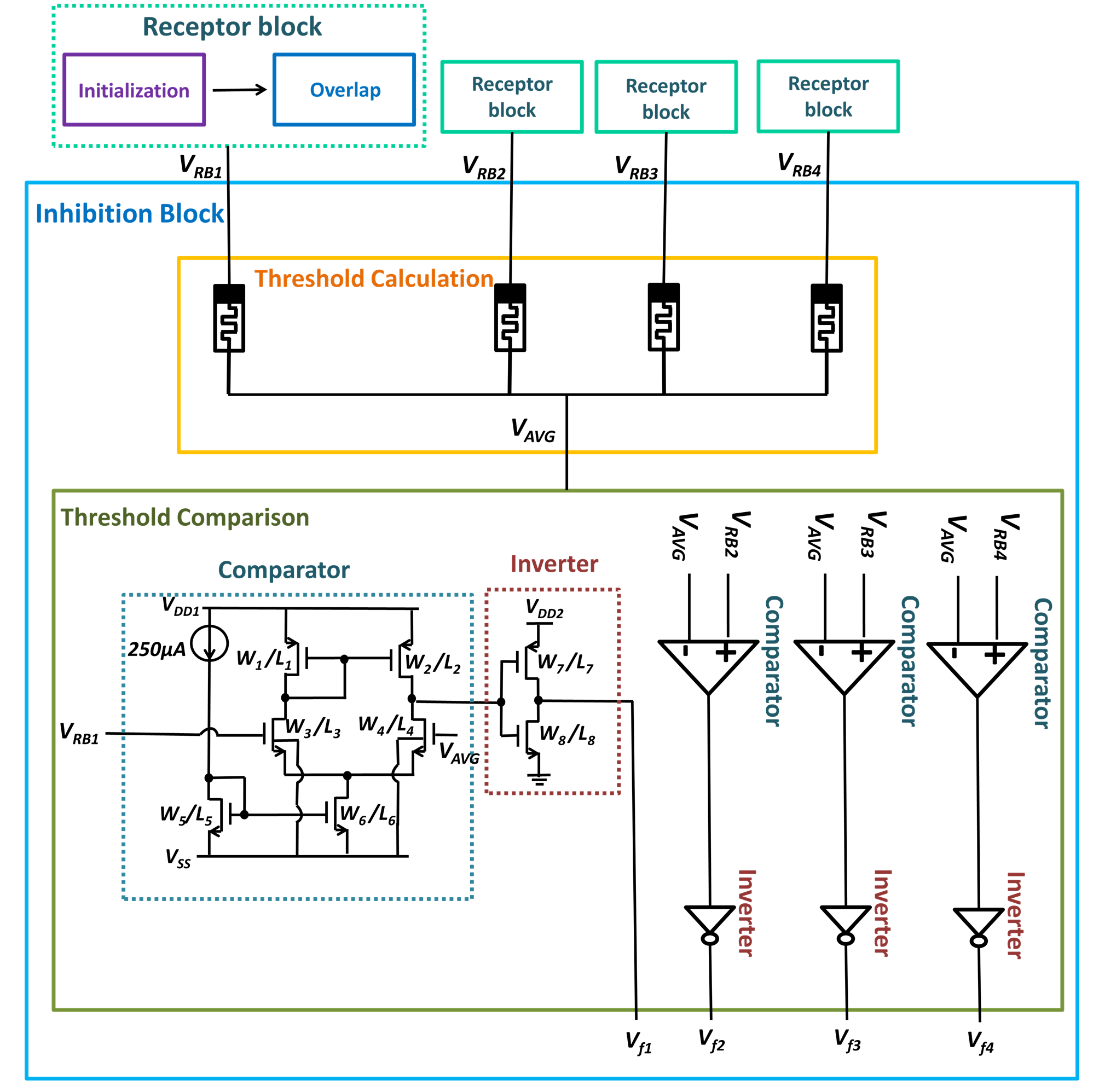}
\caption{The HTM SP structure containing receptor block and inhibition block \cite{tcad}.}
\label{fig3}
\end{figure}

As it is proven in \cite{tcad} that the modified HTM approach results in higher accuracy and reduced on-chip area and power consumption, in this work, we focus on the modified HTM algorithm and check the effect of the rule-based initialization approach for the modified HTM hardware implementation. The overall architecture of the modified HTM illustrated in Fig.\ref{fig3}. The receptor blocks correspond to the initialization and overlap calculation phases of the HTM SP and the inhibition block refers to the HTM SP inhibition phase. 

The inhibition phase consists of the memristive threshold calculation block and threshold comparison block. In the threshold calculation block, the threshold value is determined as a mean of all input overlap values, which corresponds to the modified HTM SP approach. The value of the memristors in the threshold calculation block are the same. The threshold comparison block consists of the set of comparators and inverters. Each overlap voltage, corresponding to a particular connection in the inhibition block, is followed by a single comparator and inverter. The comparator is based on the low voltage amplifier with 6 transistors and the current source. If the value of the overlap of a single column $V_{RBj}$ is greater than the overall mean of all overlaps $V_{AVG}$, the output of the comparator is low, and vise versa. To invert the output of the comparator and normalize it to a certain level, the CMOS inverter is applied. The output of the inverter in the output of the inhibition block for the particular columns, which show whether the columns columns are activated or exhibited.

\subsection{Random weight approach implementation}
\begin{figure}
\centering
\includegraphics[width=60mm]{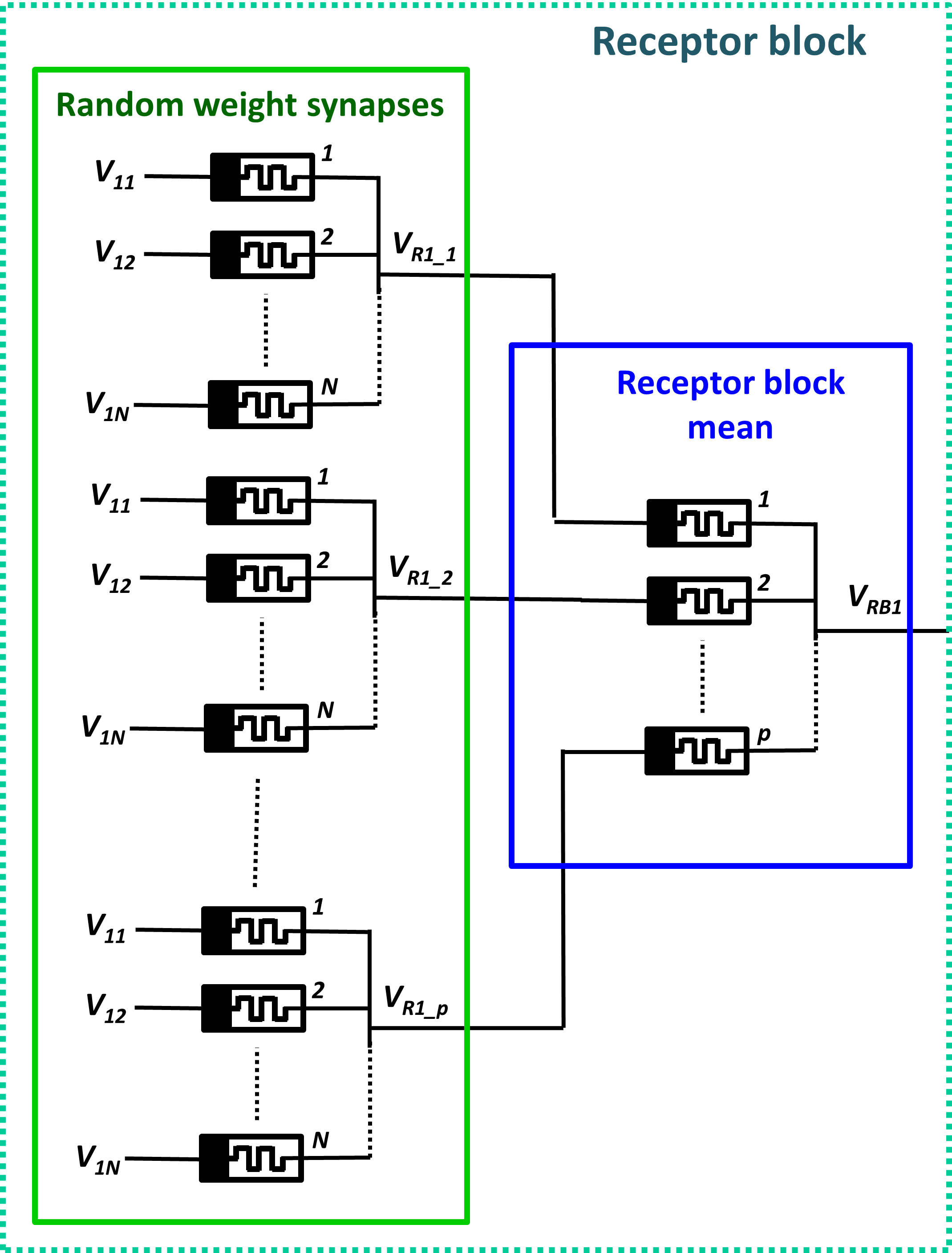}
\caption{The HTM SP receptor block structure for the random weight approach \cite{tcad}.}
\label{fig4}
\end{figure}

The difference between the traditional random weight approach and rule-based approach occurs in the receptor block of the hardware implementation of the HTM SP. The implementation of the traditional approach is illustrated in Fig. \ref{fig4}. 

The receptor block structure for the conventional HTM SP approach consists of the randomization of the weight synapses and the receptor block mean calculator. The randomization of the weights of the synapses refer to the initialization stage, where the weights are completely randomized. This is implemented with the memristive the set of the memristive mean circuits, where the resistances of the memristors are assigned randomly. Separate sets of the memristors in the block of random weight synapses refer to several random iterations to ensure that the weights are completely randomized. The receptor mean block performs the summation of all the columns for the overlap calculation. The parameter $V_{RBj}$ corresponds to the final overlap of the particular column. The tradition summation of the overlap values in the HTM SP algorithm is replaced with the mean calculation on hardware, which does not have any impact on the performance of the modified HTM SP. The resistances of the memristors in the receptor block mean are the same.

\subsection{Rule-based approach implementation}

\begin{figure}
\centering
\includegraphics[width=80mm]{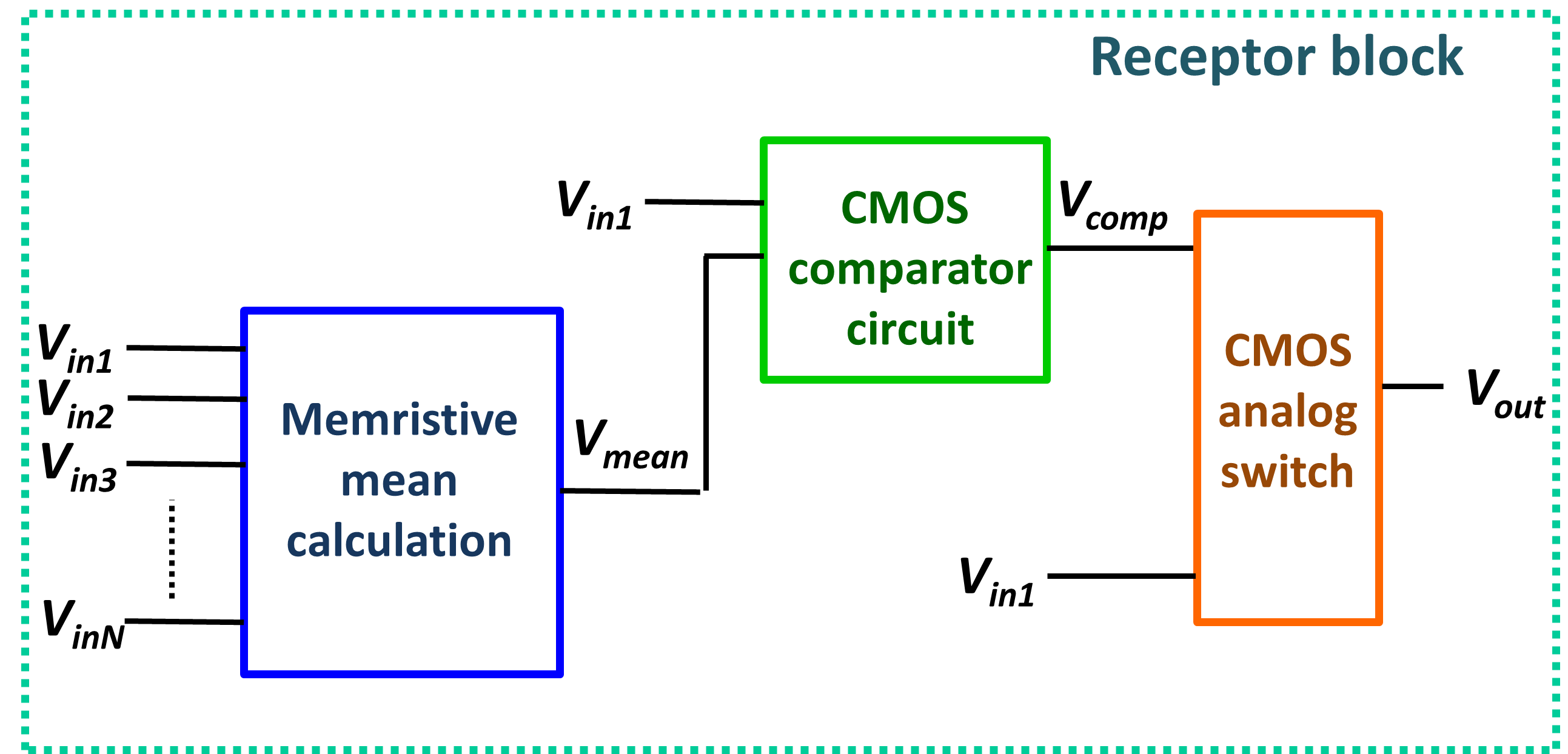}
\caption{The HTM SP receptor block structure for the rule-based approach.}
\label{fig5}
\end{figure}

In this work, we proposed the analog hardware implementation of the rule-based approach for the HTM SP. If the tradition hardware implementation of the HTM SP is based on the memristive circuits, the rule based approach is based on the CMOS circuits. The proposed receptor block is shown in Fig.\ref{fig5}. 

In the proposed architecture, the memristive mean calculation block and CMOS comparator circuit correspond to the initialization phase on the rule-based HTM SP approach. The memristive mean calculation block calculates the average of the inputs in the mean of the inputs from the set of the potential inputs. The voltage $V_{mean}$ refers to the threshold for assigning the potential inputs as connected or disconnected. The CMOS comparator circuit compares the input of the particular column with the mean of all columns from the potential inputs. If the input is greater than the threshold, the output of the comparator $V_{comp}$ is low and vice versa. The CMOS analog switch block refers to the implementation of the overlap stage of the HTM SP. The voltage $V_{out}=V_{in1}$, if the comparator output $V_{comp}$ is low, which corresponds to the case when the column is connected. The voltage $V_{out}=0$, when $V_{comp}$ is high, which means that the columns is disconnected. The voltage $V_{out}$ refers to the overlap value of the column. 

\section{System level implementation}
\label{s5}

\begin{figure*}
\centering
\includegraphics[width=170mm]{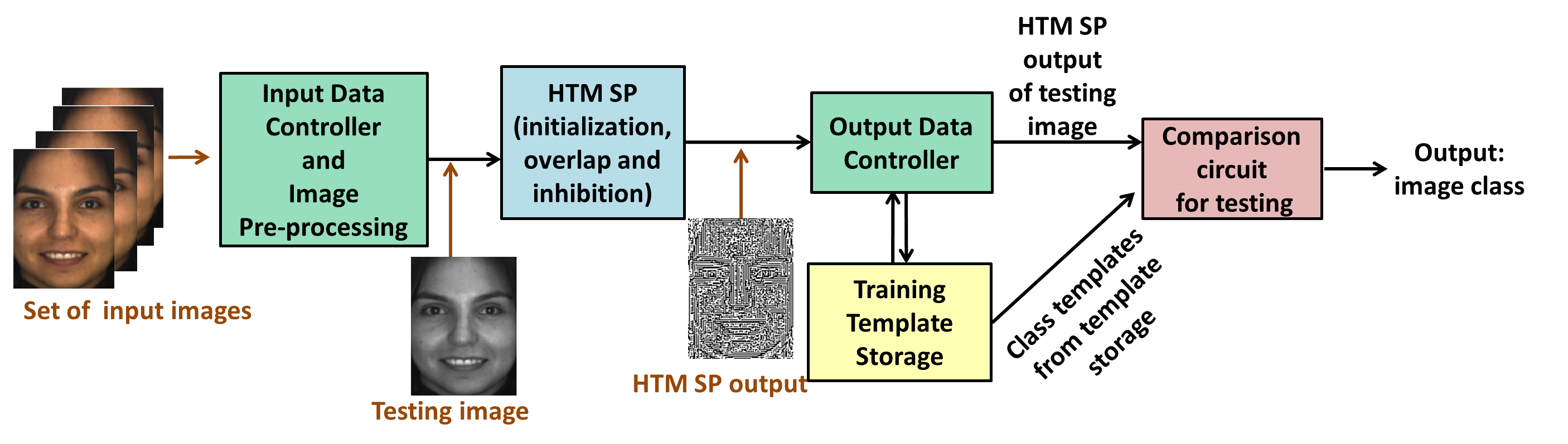}
\caption{The overall system implementation of the face recognition module with the HTM SP.}
\label{fig6}
\end{figure*}

In this work, we apply the HTM SP with two different initialization stage approach for the face recognition problem. The overall system implementation of the face recognition module with the HTM SP is illustrated in Fig.\ref{fig6}. The input RGB images are read by the image sensor and applied to the input data controller. In this stage, the sampling process occurs if it is required and the sampled images are preprocessed. In this method, we use only RGB to gray-scale conversion as a preprocessing step. In the existed HTM SP face and speech recognition systems \cite{tcad}, the standard deviation filter is applied in the preprocessing stage to improve the recognition process. However, in this work, we show the effect of the different approaches for the initialization stage; therefore, we remove the filtering stage to obtain the actual results from the HTM SP. 

After the controller, the image is applied to the HTM SP stage, which performs the encoding of the image and outputs the sparse binary image with the preserved important image features. The output data controller controls, where the images are directed in the training and testing stages. In the training stage, the output from the HTM SP is preserved in the training template storage. The training continues until all image class templates are preserved. In the testing stage, the output data controller directs the images into the comparison circuit. The comparison circuit can be implemented as a memristive pattern matcher, which compares all templates form the training template storage with the current input image. Finally, the image class is determined.

The algorithmic implementation of the face recognition system approach is shown in Algorithm \ref{algor1} in Appendix.

\section{Results}
\label{s6}

\subsection{System level simulation}

\begin{figure}[!ht]
    \centering
    \subfigure[]
    	{
     \centering	\includegraphics[width=0.08\textwidth]{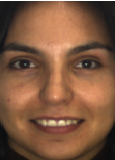}
    	\label{f11}
		}
	\subfigure[]
		{
     \centering	\includegraphics[width=0.08\textwidth]{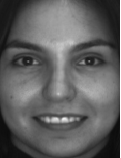}
    	\label{f22}
		}
     \subfigure[]
		{
         \centering
\includegraphics[width=0.08\textwidth]{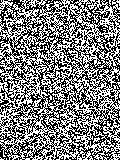}
    	\label{f6}
}  
   \subfigure[]
		{
         \centering
\includegraphics[width=0.08\textwidth]{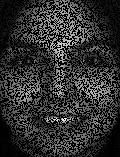}
    	\label{f7}
}  
  \subfigure[]
		{
         \centering
\includegraphics[width=0.08\textwidth]{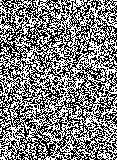}
    	\label{f8}
}  
    \caption{Simulation results for the random weight approach: (a) input image, (b) grayscale image, (c) binary weights, (d) HTM SP overlap output and (e) HTM SP inhibition output.}
     \label{fig7a}
\end{figure}

\begin{figure}[!ht]
 \label{TMresultss}
    \centering
    \subfigure[]
    	{
     \centering	\includegraphics[width=0.08\textwidth]{Figures/f1}
    	\label{f1}
		}
	\subfigure[]
		{
     \centering	\includegraphics[width=0.08\textwidth]{Figures/f2}
    	\label{f2}
		}
       \subfigure[]
		{
         \centering
\includegraphics[width=0.08\textwidth]{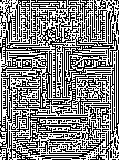}
    	\label{f3}
}  
   \subfigure[]
		{
         \centering
\includegraphics[width=0.08\textwidth]{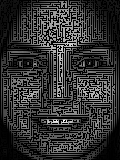}
    	\label{f4}
}  
  \subfigure[]
		{
         \centering
\includegraphics[width=0.08\textwidth]{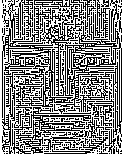}
    	\label{f5}
}  
    \caption{Simulation results for the rule-based approach with 2 inputs in the receptor region: (a) input image, (b) grayscale image, (c) binary weights, (d) HTM SP overlap output and (e) HTM SP inhibition output.}
     \label{fig7b}
\end{figure}

The experiments for the system level simulation were performed in MATLAB for 3 different databases: AR, ORL and YALE. The AR database contains 100 classes of faces with 26 face images per class with various natural variation and occlusions \cite{martinez1998ar}. The ORL database includes 40 classes with 10 image per class with occlusions, scale variations and rotations \cite{46}. The YALE database contains 15 classes with 11 images per class including different facial expressions and natural variabilities \cite{47}. For the experiments in this work, 50\% of the images with used for training and the other 50\% for testing. The exemplar images for the random weight approach are shown in Fig.\ref{fig7a}, and for the rule-based approach in Fig.\ref{fig7b}. 

\begin{figure}[!ht]
    \centering
    \subfigure[]
    	{
     \centering	\includegraphics[width=0.45\textwidth]{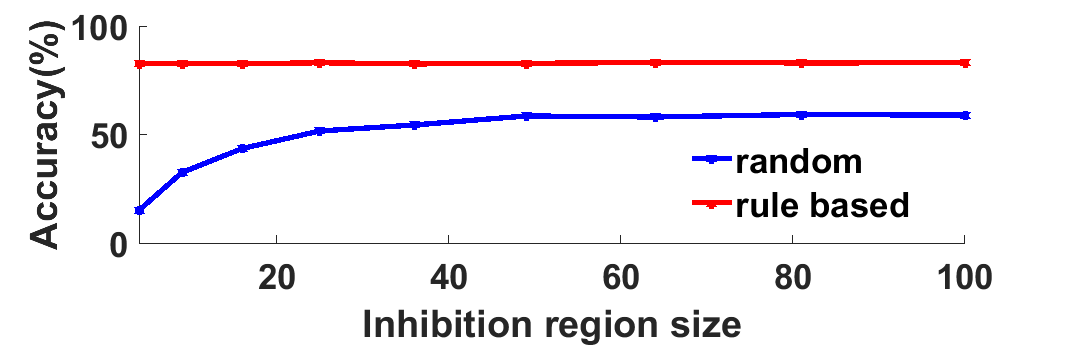}
    	\label{f8a}
		}
	\subfigure[]
		{
     \centering	\includegraphics[width=0.45\textwidth]{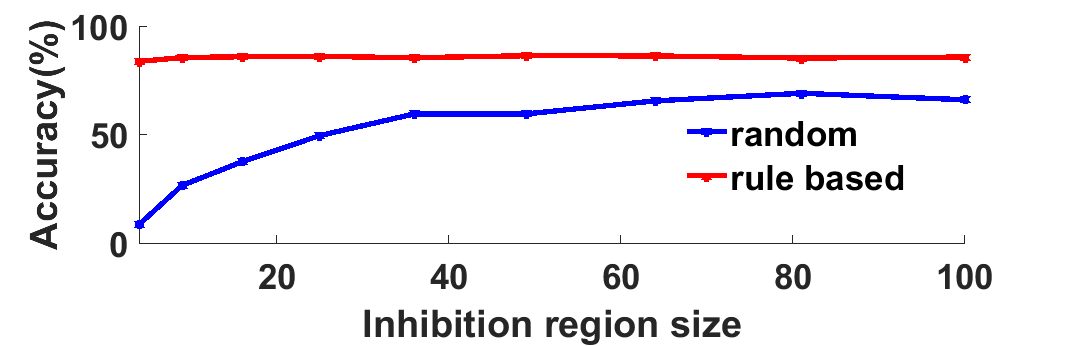}
    	\label{f8b}
		}
       \subfigure[]
		{
         \centering
\includegraphics[width=0.45\textwidth]{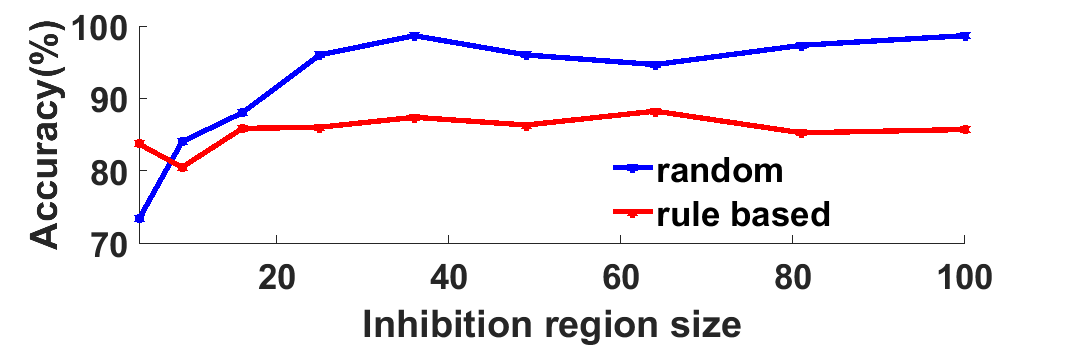}
    	\label{f8c}
}  
  
    \caption{Simulation results for the face recognition for two methods for different databases: (a) AR, (b) ORL and (c) YALE.}
     \label{fig8}
\end{figure}

The recognition accuracy of the random weight and rule base approaches with the variation of the size of the inhibition region is shown in Fig.\ref{fig8}. Fig.\ref{f8a} illustrates the simulation results for AR database, Fig.\ref{f8b} for ORL database and Fig.\ref{f8c} for YALE database. The rule-base approach improves face recognition accuracy for AR and ORL databases. However, for the YALE database, the recognition accuracy is decreased. This can be explained by the small number of classes and face samples in the YALE database. The average and maximum recognition accuracies for two approaches are compared in Table \ref{t1}.

\begin{table}[]
\centering
\caption{The average and maximum recognition accuracies for different databases for traditional random weight and proposed rule based approaches.}
\label{t1}
\begin{tabular}{|l|l|c|c|c|c|}
\hline
\multicolumn{2}{|l|}{\multirow{2}{*}{\textbf{Dataset}}} & \multicolumn{2}{c|}{\textbf{\begin{tabular}[c]{@{}c@{}}Random weight\\  approach\end{tabular}}}                       & \multicolumn{2}{c|}{\textbf{\begin{tabular}[c]{@{}c@{}}Rule based \\ approach\end{tabular}}}                           \\ \cline{3-6} 
\multicolumn{2}{|l|}{}                                   & \begin{tabular}[c]{@{}c@{}}mean\\  accuracy\end{tabular} & \begin{tabular}[c]{@{}c@{}}maximum\\ accuracy\end{tabular} & \begin{tabular}[c]{@{}c@{}}mean \\ accuracy\end{tabular} & \begin{tabular}[c]{@{}c@{}}maximum\\  accuracy\end{tabular} \\ \hline
\multicolumn{2}{|l|}{AR}                                 & 47.992 \%                                               & 59.231 \%                                                 & 82.855\%                                                & 83.231\%                                                   \\ \hline
\multicolumn{2}{|l|}{YALE}                               & 91.852 \%                                               & 98.667 \%                                                 & 85.538 \%                                               & 86.308 \%                                                  \\ \hline
\multicolumn{2}{|l|}{ORL}                                & 49.056 \%                                               & 69.000 \%                                                 & 85.5389 \%                                               & 86.308 \%                                                  \\ \hline
\end{tabular}
\end{table}

\subsection{Analog hardware simulation}

\begin{figure}[!ht]
    \centering
    \subfigure[]
    	{
     \centering	\includegraphics[width=0.4\textwidth]{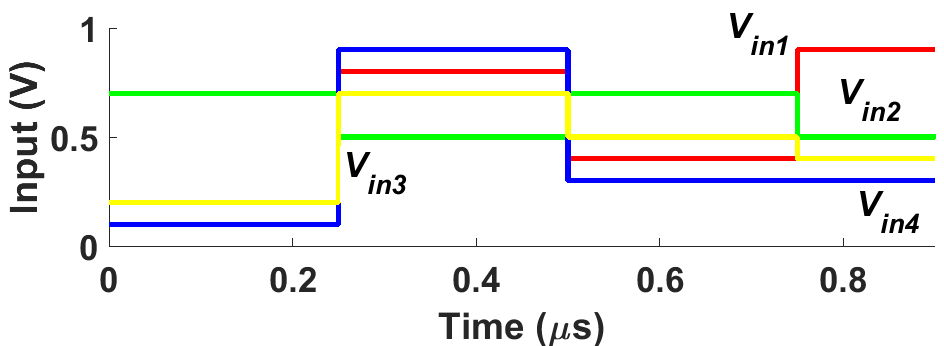}
    	\label{fig9a}
		}
	\subfigure[]
		{
     \centering	\includegraphics[width=0.4\textwidth]{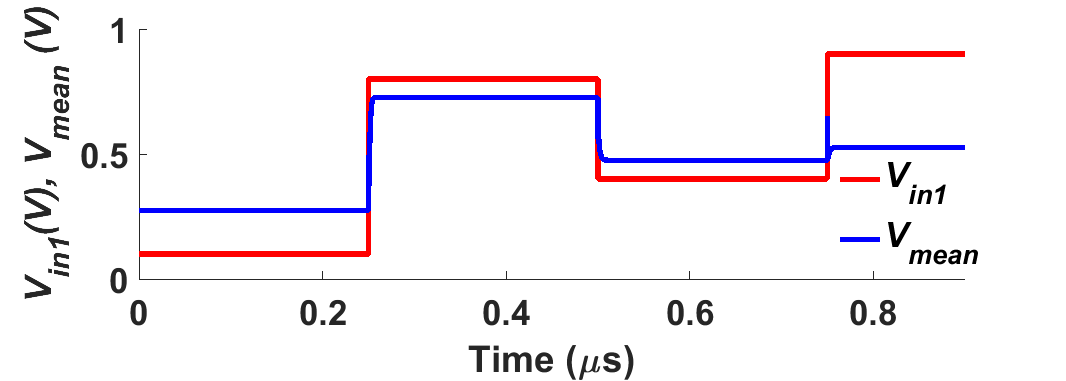}
    	\label{fig9b}
		}
   
    \subfigure[]
		{
         \centering
\includegraphics[width=0.4\textwidth]{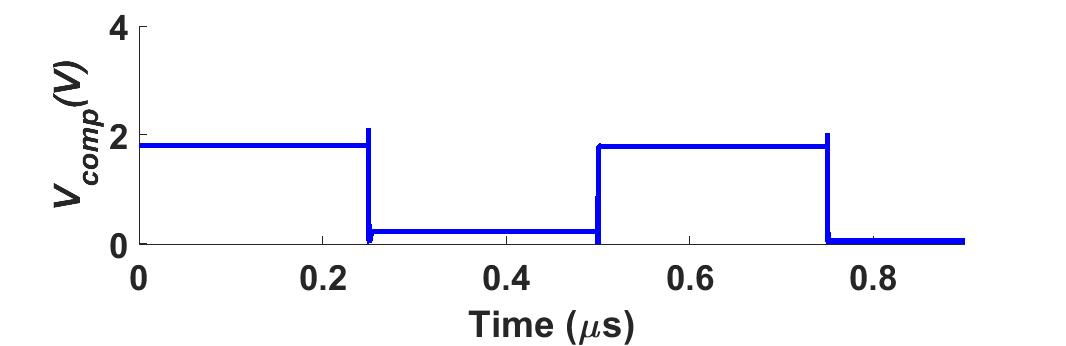}
    	\label{fig9c}
}  
   \subfigure[]
		{
         \centering
\includegraphics[width=0.4\textwidth]{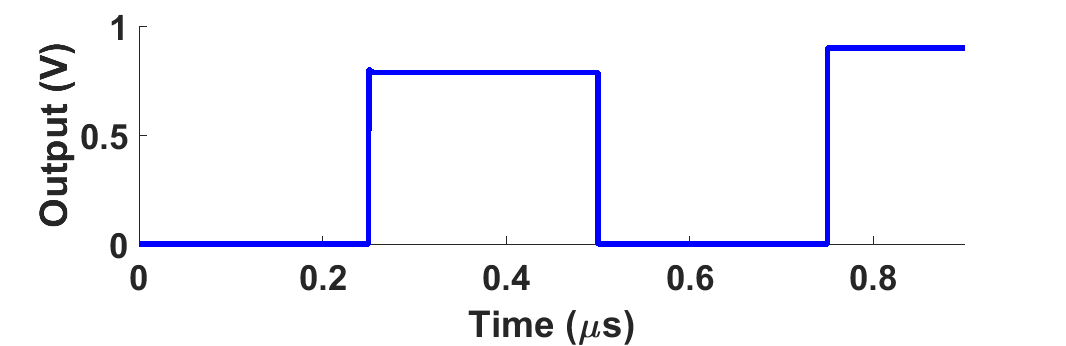}
    	\label{fig9d}
}  
    \caption{Timing diagram for the proposed receptor block for the rule-based HTM approach: (a) inputs from the neighborhood, (b) main input and mean of the inputs, (c) comparator output and (d) receptor block output.}
     \label{fig9}
\end{figure}

The simulation of the proposed rule-based approach was performed in SPICE for TSMC 180nm CMOS technology. Fig.\ref{fig9} illustrates the timing diagram for the proposed rule-based receptor block, shown in Fig.\ref{fig5}. Fig.\ref{fig9a} shows the inputs in the receptor block. Fig.\ref{fig9b} illustrates the main input and the total mean of all the inputs. This main input is compared with the mean in the following stages. Fig.\ref{fig9c} shows the comparator circuit output and Fig.\ref{fig9d} illustrates the final output of a single receptor block. 

Table \ref{t2} compares the on-chip area and power dissipation for random weight and rule-based approaches.

\begin{table}[]
\centering
\caption{Comparison of the random weight and rule base approaches in terms of the on-chip area and maximum power consumption of a single receptor block.}
\label{t2}
\begin{tabular}{|l|c|c|}
\hline
\textbf{Approach}      & \textbf{\begin{tabular}[c]{@{}c@{}}On-chip \\ area\end{tabular}} & \textbf{\begin{tabular}[c]{@{}c@{}}Power\\  dissipation\end{tabular}} \\ \hline
Random weight approach & $0.125\mu m^2$                                                   & $42.92 pW$                                                            \\ \hline
Rule-based approach    & $13.31\mu m^2$                                                   & $135\mu W$                                                            \\ \hline
\end{tabular}
\end{table}

\section{Discussion}
\label{s7}

As it was illustrated in Section \ref{s6}, the proposed rule-based approach outperforms the traditional HTM random weight approach. This can be explained by the fact that the rule-based approach that draws the correlation between the HTM SP weights to the input space. The main goal of the HTM SP is to create the SDR from the input. However, the facial images contain the natural sparseness. The rule based approach ensures the preservation of this natural sparseness of the images, which results in the increase of the recognition accuracy. In addition, this allows to preserve the structural information from the images, such as edges.

The hardware implementation of the rule-based approach required larger on-chip area and power consumption, comparing to the traditional random weight method. However, to achieve high recognition accuracy in the rule-based approach, the image filtering stage is not required, which is performed on the separate software. Moreover, the rule-based approach does not require the programming of the memristors to the random weights, which can be achieved combining either software-based or mixed-signal random number generation approach. The programming of the memristors requires additional time and reduces the processing speed. Also, the high accuracy of the rule-based approach result allows to remove the learning phase from the HTM SP, which can be implemented using digital or analog circuits and requires a significant amount of extra power and on-chip area \cite{tcad}.

\section{Conclusion}
\label{s8}
In this paper, the hardware implementation of a rule-based approach for the initialization phase of the HTM SP has been proposed. The proposed rule-based approach allows to achieve significant increase in recognition accuracy. The maximum accuracy is approximately $86 \%$, which is equivalent to the processing of the HTM SP with the learning phase. The on-chip area and power requirements to implement the rule-based initialization phase of the HTM SP are $13.31\mu m^2$ and $135\mu W$ for a single receptor block, respectively.  

\section*{Appendix}
\label{app}
In Algorithm \ref{algor1}, line 2 refers to the preprocessing stage, lines 3-17 refer to the HTM SP processing, lines 18-19 correspond to the training phase and lines 20-22 shows the testing (recognition) phase.

\begin{algorithm}

\caption{System level implementation of HTM}
\label{algor1}
\begin{algorithmic}[1]

\State $Define$ $neighborhood$ $N$
\State $x = grayscale(x)$
\Comment{\textbf{PRE-PROCESSING}}

\For {$p$ inhibition regions}
\Comment{\textbf{HTM SP}}
\For {$k$ image blocks} 
 \ForAll{$i \in W$}
 \If{$x(i) \geqslant mean(x(i) \in N)$} 
 \State $W(i) = 1$
 \Else
 \State $W(i) = 0$
 \EndIf
 \EndFor
\State $image.block (j) =W(j)\times image.block(j)$ 
 \EndFor
\State$threshold.block = mean (y.image.blocks)$
\For{$y$ image blocks}
\If{$image.block(y) > threshold.block$} 
\State $inhibition.region(y) = 1$
\Else
\State $inhibition.region(y) = 0$
\EndIf
\EndFor
\State $x(p) = inhibition.region(p)$
\EndFor
\If{training phase}
\State $Store$ $image$ $to$ $the$ $training$ $template$
\ElsIf{testing phase}
\State $Compare$ $image$ $to$ $all$ $atored$ $templates$
\State $Determine$ $image$ $class$
\EndIf
\end{algorithmic}
\end{algorithm}



\bibliographystyle{spmpsci}      

\bibliography{ref.bib}   


\end{document}